\documentclass[english,10pt,amsmath,amssymb,a4paper]{article}
\usepackage{graphicx,babel,geometry,amsmath,amssymb}
\usepackage[T1]{fontenc}
\usepackage[latin9]{inputenc}
\begin{document}

\title{Origins of Binary Gene Expression in Post-Transcriptional Regulation by MicroRNAs}
\author{Indrani Bose\thanks{indrani@bosemain.boseinst.ac.in} \, and Sayantari Ghosh  \\
        Department of Physics, Bose Institute, 93/1, A. P. C. Road, Kolkata - 700009, India
        }
\date{}

\maketitle

\begin{abstract}
MicroRNA-mediated regulation of gene expression is characterised by some distinctive features that set it apart 
from unregulated and transcription factor-regulated gene expression. Recently, a mathematical 
model has been proposed to describe the dynamics of post-transcriptional regulation by microRNAs. 
The model explains the observations made in single cell experiments quite well. In this paper, 
we introduce some additional features into the model and consider two specific cases. In the first case, a
 non-cooperative positive feedback loop is included in the transcriptional regulation of the
 target gene expression. In the second case, a stochastic version of the original model is considered in 
which there are random transitions between the inactive and active expression states of the gene. In the 
first case we show that bistability is possible in a parameter regime, due to the
 presence of a non-linear protein decay term in the gene expression dynamics. In the second
 case, we derive the conditions for obtaining stochastic binary gene expression. We find that
 this type of gene expression is more favourable in the case of regulation by microRNAs as compared to 
 the case of unregulated gene expression. The theoretical predictions relating to binary gene
 expression are experimentally testable.
\end{abstract}

\section {Introduction}
MicroRNAs (miRNAs) are a class of small, noncoding RNAs which regulate gene expression in
prokaryotes and eukaryotes at the post-transcriptional level. They play critical roles in a 
number of cellular processes, such as stress response, developmental transitions, differentiation,
apoptosis etc. \cite{inui, flynt, leung}. The mechanisms of regulation by small RNAs differ
in specific features in prokaryotes and eukaryotes. These are, however, based on the common 
principle of the regulatory RNA base-pairing with the messenger RNA (mRNA) of the target gene
inhibiting translation and/or promoting mRNA degradation. Various possibilities have been
 suggested for the fate of the inactive complex of the regulatory RNA and the target mRNA 
 once it is formed \cite{mukherji, which, levine,levine2}: (i) the complex has a finite lifetime
 followed by dissociation into its free components, (ii) the regulatory RNA co-degrades with
 the target mRNA at the same or different rates and (iii) only the target mRNA degrades with
the regulatory RNA becoming free for further activity \cite{mukherji, which, levine,levine2}. 
 A number of functional features of RNA-regulated gene expression have been identified so far 
 \cite{mukherji, levine,levine2}. One prominent feature is that of a threshold linear mode
 of action, in which the target gene protein synthesis is highly repressed below a threshold 
 level of target mRNA production and activated in a linear fashion once the threshold is 
 crossed. Other significant characteristics include: suppression of protein fluctuations in 
 the form of translational bursts, rapid response times and filtering of transient signals
 \cite{levine2}, sharpening of spatial expression patterns \cite{levine3} and prioritisation 
 of the expression of target genes in the case of a single RNA regulating the expression
 of multiple genes \cite {mitarai}.

\begin{figure}
\centering{}
\includegraphics[scale=0.4]{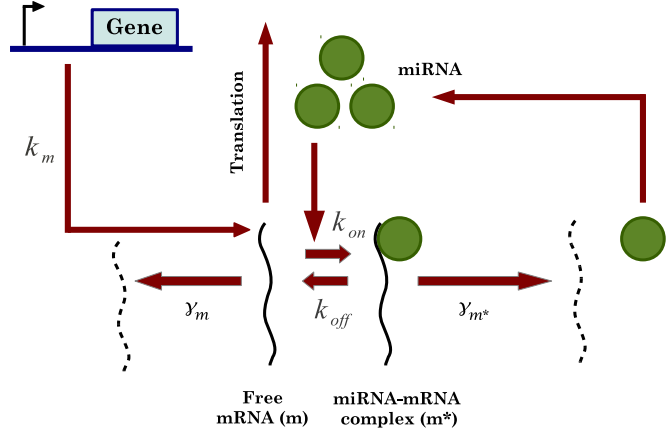}
\caption{A schematic diagram describing the biochemical events involved in miRNA-regulated gene expression. The target mRNA is synthesised and degraded with
 rate constants $k_{m}$ and $\gamma_{m}$ respectively. The rate constants $k_{on}$ and $k_{off}$ describe the formation and dissociation of the bound complex $m^{*}$ 
 of free mRNA $m$ and the miRNA. The bound miRNA degrades the mRNA with rate constant $\gamma_{m^{*}}$.}
\label{Fig_scheme}
\end{figure}

Recently, a mathematical model has been proposed \cite{mukherji} to describe the biochemical 
 interactions and kinetics of miRNA-mediated regulation of gene expression. The model has been 
 experimentally validated in single cell measurements using quantitative fluorescence 
 microscopy and flow cytometry. The experiments clearly demonstrate the high repression 
 of target protein synthesis below a threshold level of target mRNA production, and a sensitive 
 response above the threshold. A major finding is that the strength of the repression 
 below the threshold has considerable cell-to-cell variation in a population of genetically-identical 
cells. This indicates that stochasticity may have non-trivial consequences in miRNA-mediated 
 regulation of gene expression. Figure \ref{Fig_scheme} shows a sketch of the different biochemical processes
 involved in the miRNA-mediated regulation of gene expression, and which form the basis of the
 mathematical model proposed in \cite{mukherji}. The mRNA, as shown in Fig. \ref {Fig_scheme}, is either free
 ($m$) or is part of a miRNA-mRNA complex ($m^{*}$). The target mRNA is transcribed with rate constant
 $k_{m}$ and has a natural degradation rate constant $\gamma_{m}$. The rate constants $k_{on}$ and $k_{off}$
 are associated with the formation and dissociation of the bound complex of mRNA and miRNA. The 
bound miRNA becomes free either by unbinding the target mRNA (rate constant $k_{off})$ or by 
degrading the mRNA (rate constant $\gamma_{m^{*}}$). The model incorporates the important feature
 of molecular titration similar to the protein-protein titration analysed earlier \cite{buchler}.
 Protein sequestration occurs when a repressor protein binds an active protein thus forming an
 inactive complex. As shown in \cite{buchler}, the regulatory mechanism, termed molecular titration,
 can generate ultrasensitive input-output responses. In the case of miRNA-mediated regulation of gene
 expression, the inactive complex is that of mRNA and miRNA. Molecular titration is responsible for
 the observed sensitive dependence of protein expression on target mRNA input around a threshold level
 of target mRNA production.

In this paper, we analyse the dynamics of the model, the key biochemical processes
 of which are shown in Fig. \ref {Fig_scheme}, from two different perspectives. In the first case,
 an additional process representing the autoactivation of the target gene expression by the target
 protein is included. In the second case, a simple stochastic version of the model of Fig. \ref {Fig_scheme}
 is considered. In this model, the only stochasticity is associated with the random transitions between the
 inactive and active expression states of the gene. In the inactive state, there is no transcriptional activity, i.e.,
 mRNA synthesis. Transcription is only initiated in the active state of the gene. The degradation of the mRNA
 occurs in both the inactive and active states of the gene. We show that binary gene expression is a possible
 outcome in both of the scenarios described above. In the deterministic case, binary gene expression implies bistability,
 i.e., the coexistence of two stable expression states. In the stochastic case, the distribution of mRNA levels is
 bimodal, i.e., has two prominent peaks. In the following, we analyse the deterministic and stochastic
 models to investigate the origins of binary gene expression. 

\section {Deterministic Model} 
The total concentration of miRNA is assumed  to be constant, since the miRNA turnover is slow compared to the timescale
 of gene expression \cite {mukherji}. The following set of differential equations describe the dynamics of the model.

\begin{equation}
\centering{}
\frac{dm}{dt}=k_{m}+\frac{\beta_{m}p}{K+p}-k_{on\,}m.mi+k_{off}\, m^{*}-\gamma_{m\,}m
\label{Eq_dmdt}
\end{equation}

\begin{equation}
\centering{}
\frac{dm^{*}}{dt}=k_{on}\, m.mi-k_{off}\, m^{*}-\gamma_{m^{*\,}}m^{*}
\label{Eq_dm*dt}
\end{equation}

\begin{equation}
\centering{}
\frac{dp}{dt}=k_{p}\, m-\gamma_{p\,}p
\label{Eq_dpdt}
\end{equation} The conservation condition for miRNA is 

\begin{equation}
\centering{}
mi_{T}=mi+m^{*}
\label{Eq_mi}
\end{equation} In the equations above, $mi$ and $mi_T$ represent the free and total miRNA concentrations. The second term on the r.h.s. of Eq. (1) represents
 the autoactivation of the target gene expression. The rate constant $\beta_{m}$ represents the maximum rate of mRNA
synthesis due to autoactivation, with $K$ denoting the equilibrium
dissociation constant for the binding of the regulatory protein at
the promoter region of the target gene. The rate constants $k_{p}$
and $\gamma_{p}$ correspond to protein synthesis and degradation,
respectively, with $p$ being the protein concentration. In the steady state, $\frac{dm}{dt}=0$, $\frac{dm^{*}}{dt}=0$ and $\frac{dp}{dt}=0$
 and one obtains the equations

\begin{equation}
\centering{}
k_{m}+\frac{\beta_{m}p}{K+p}-k_{on\,}m.mi+k_{off\,}m^{*}-\gamma_{m\,}m=0, \: \: \:  \gamma^{*}=\gamma_{m^{*}}.mi_{T}
\label{Eq_ss1}
\end{equation}

\begin{equation}
\centering{}
m^{*}=\frac{m.mi_{T}}{m+\lambda}, \: \:  \lambda=\frac{k_{off}+\gamma_{m^{*}}}{k_{on}}
\label{Eq_ss2}
\end{equation}

\begin{equation}
\centering{}
p=\frac{k_{p}}{\gamma_{p}}m
\label{Eq_ss3}
\end{equation}From these equations, the steady state protein concentration satisfies the equation 

\begin{equation}
\centering{}
p_{st}=k_{\delta}+\frac{\alpha \, p_{st}}{K+p_{st}}-\frac{\phi \, p_{st}}{p_{st}+\lambda_{\alpha}}
\label{Eq_sss}
\end{equation}with $k_\delta=\frac{k_{m}}{\gamma_{\alpha}}$, $\alpha=\frac{\beta_{m}}{\gamma_{\alpha}}$, 
$\phi=\frac{\gamma^{*}}{\gamma_{\alpha}}$, $\lambda_{\alpha}=\lambda \: \frac{k_{p}}{\gamma_{p}}$ and $\gamma_{\alpha}=\frac{\gamma_{m}\gamma_{p}}{k_{p}}$.

\begin{figure}
\centering{}
\includegraphics[scale=0.92]{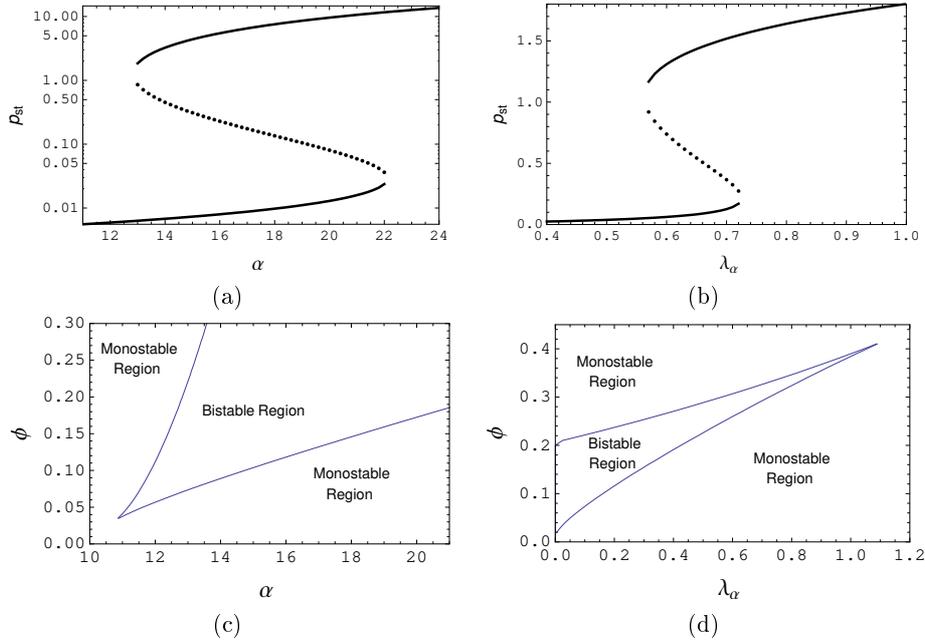}
\caption{(a) Steady state protein concentration $p_{st}$ versus $\alpha$. The parameter values are $k_{\delta}=0.01$, $K=10$, $\phi=0.2$ and $\lambda_{\alpha}=0.1$.
(b)  Steady state protein concentration $p_{st}$ versus $\lambda_{\alpha}$. The parameter values are $k_{\delta}=0.01$, $K=10$, $\phi=0.3$ and $\alpha=13$. 
(c) Phase diagram in the $\phi-\alpha$ plane showing a region of bistability between two regions of monostability. A continuous path between the two monostable regions 
is also possible. The parameter values are $k_{\delta}=0.01$, $K=10$ and $\lambda_{\alpha}=0.1$. 
(d) Phase diagram in the $\phi-\lambda_{\alpha}$ plane showing a region 
of bistability between two regions of monostability. The parameter values are $k_{\delta}=0.01$, $K=10$ and $\alpha=13$.}
\label{Fig_hyst}
\end{figure}
\noindent Eq. (\ref{Eq_sss}) has two stable steady state solutions, i.e., bistability in specific parameter regions. Figure 2(a)
shows a plot of the steady
state protein concentration, $p_{st}$, as a function of the parameter
$\alpha$. The stable steady states are represented by solid lines
whereas the dotted line describes the branch of unstable steady states.
The plot exhibits hysteresis since the discontinuous transitions from
the lower to the upper branch, and from the upper to the lower branch,
occur at different values of the parameter $\alpha$, the so-called
 bifurcation points of the associated dynamics. Bistability constitutes
 a universal theme in several cell biological processes \cite{smits, pomer}
 and signifies that a cell has a choice between two stable expression states
 for the same parameter values. The choice of the specific stable state
 depends on the previous history of the system. Figure 2(b) exhibits a plot of $p_{st}$ versus the parameter $\lambda_{\alpha}$. Figure 2(c) shows a phase
diagram in the $\phi-\alpha$ plane in which a region of bistability
separates two regions of monostability, corresponding to low and high
expression states, respectively. Figure 2(d) shows a similar phase diagram in the $\phi-\lambda_{\alpha}$ plane. 
Bistability, in general, is an outcome
of dynamics involving positive feedback and sufficient nonlinearity.
The latter condition is usually achieved via the binding of the regulatory
protein molecules at multiple sites of the promoter region of the
gene, or when the regulatory proteins form multimers, such as dimers and
tetramers, which then bind specific regions of the DNA (cooperativity
in regulation) \cite{ferel, mitro}. In the case of transcriptional regulation, 
Tan et al. \cite {tan} have proposed a new mechanism by which a noncooperative positive
 feedback loop combined with a nonlinear protein decay term are sufficient to generate
 bistability. This novel type of bistability was demonstrated in the operation of
 a synthetic gene circuit. The circuit contains a single autoregulatory positive feedback 
loop in which the protein product X of a gene promotes its own synthesis in a noncooperative fashion. The protein 
decay rate is a sum of two terms, the natural degradation rate and the dilution rate due to cell growth. In the 
synthetic circuit, the production of X inhibits cell growth so that the dilution rate of X and hence the protein 
decay rate are reduced. This gives rise to an effective positive feedback loop, since an increased synthesis of X proteins 
leads to a greater accumulation of the proteins, which in turn activates further protein synthesis.
 The combination of two positive feedback loops gives rise to bistability in a parameter regime in the absence of cooperativity. 
A related study by Klumpp et al. \cite {klump} has also demonstrated the generation of a positive feedback loop due to cell growth inhibition by a protein. The nonlinear 
protein decay term in these cases (`emergent bistability') has the same form as in Eq. (8). The origin of the nonlinearity is, however, different in the case of 
the post-transcriptional, i.e., miRNA-mediated regulation of gene expression. The effect of a transcriptional positive feedback loop on miRNA-regulated
 gene expression has been investigated in a number of earlier studies \cite {zhad}. Specifically, bistability has been demonstrated in cases where
 the transcriptional positive feedback loop involves some form of cooperativity. Our present study establishes a new origin of bistability in
 miRNA-regulated gene expression, based on a non-cooperative transcriptional positive feedback loop and a nonlinear protein decay term. 
Since our model closely follows the experimentally-tested model proposed in Ref. \cite {mukherji}, the inclusion of a transcriptional positive feedback loop in the original gene circuit could provide an experimental test 
of the new mechanism of bistability suggested by us.

\section{Stochastic Model}

The model is a generalisation, incorporating miRNA-mediated regulation
of gene expression, of an earlier stochastic model \cite{karma} in which
the effect of stochasticity on unregulated/transcriptional factor-regulated gene expression was considered.
More specifically, the condition for obtaining stochastic binary gene
expression in terms of two gene expression parameters, $r_{1}=\frac{k_{a}}{\gamma_{m}}$
and $r_{2}=\frac{k_{d}}{\gamma_{m}}$ was derived in the previous study. The rate constants $k_{a}$
and $k_{d}$ are the activation and deactivation rate constants for
transitions between the inactive and active states of the gene. The
rate constant $\gamma_{m}$ is the degradation rate constant of the gene
expression product, which could be either mRNA or protein. In the present
study, we determine the steady state distribution of mRNA levels.
We assume that the concentration $m^{*}$ of the mRNA-miRNA complex attains
its steady state value at an earlier time point than the concentration $m$ of free
mRNA. As already mentioned, the only stochasticity
in the model considered here is associated with the random transitions
between the inactive and active states of the gene. In each state
of the gene, the mRNA concentration evolves according to the equation,

\begin{figure}
\centering{}
\includegraphics[scale=0.61]{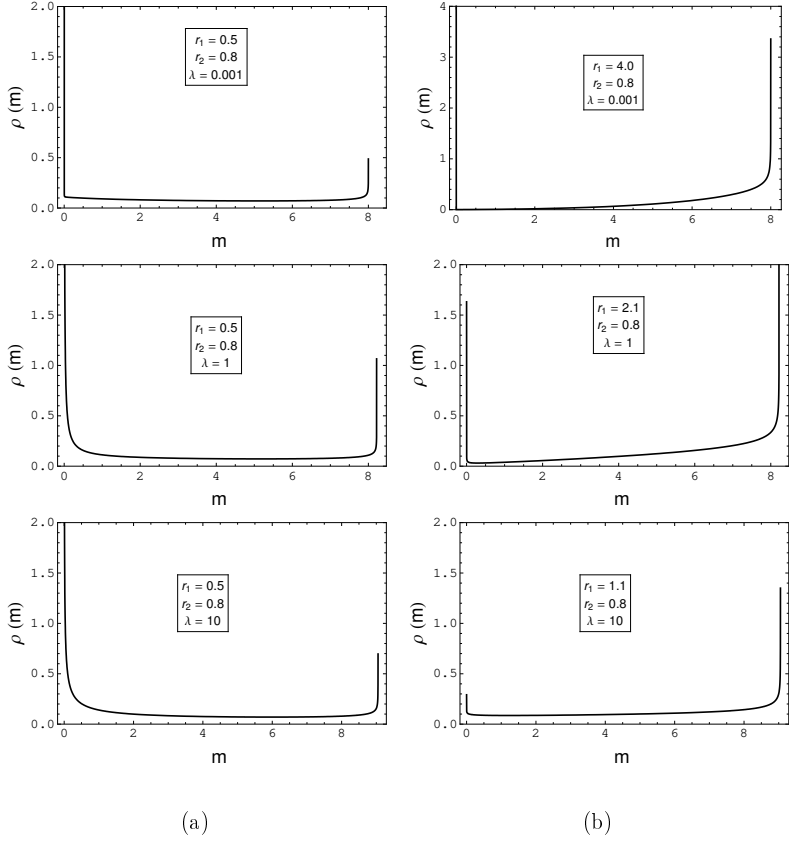}
\caption{Plots of $\rho(m)$ versus $m$ (Eq. (14)) in the steady state ($a_{1}<0$ and $a_{4}+a_{5}<0$). The values of $r_{1}$, $r_{2}$ and $\lambda$ are shown in individual 
boxes. The other parameter values are  $\gamma_{new}=0.1$, $\gamma_{new}^{*}=0.2$. (a) $r_{1}<1$, $r_{2}<1$,  (b) $r_{1}>1$, $r_{2}<1$. For a wide range of $\lambda$ 
values, binary gene expression is obtained in each case. }
\label{Fig_extra}
\end{figure}

\begin{equation}
\frac{dm}{dt}=k_{m}z-\frac{\gamma^{*}\,m}{m+\lambda}-\gamma_{m\,}m
\label{Eq_fmz}
\end{equation}Eq. (\ref {Eq_fmz}) is obtained from Eq. (1) by ignoring the positive feedback term and substituting the steady state concentration of $m^{*}$ (Eq. (6)). 
The variable $z=1\:(0)$ when the gene is in the active (inactive)
state and switches values with stochastic rate constants $k_{a}$ ($0 \rightarrow 1$) and $k_{d}$ ($1 \rightarrow 0$) . Let $\rho_{j}(m,t)$ $(j=0,1)$ be the probability density function
when $z=j$ with the total probability density function $\rho(m,t)=\rho_{0}(m,t)+\rho_{1}(m,t)$.
The rate equations for the probability density functions are given
by

\begin{equation}
\frac{\partial\rho_{0}(m,t)}{\partial t}=-\frac{\partial}{\partial m}\{(-\frac{\gamma^{*}m}{m+\lambda}-\gamma_{m}\, m)\rho_{0}(m,t)\}+k_{d}\,\rho_{1}(m,t)-k_{a\,}\rho_{0}(m,t)
\end{equation}

\begin{equation}
\frac{\partial\rho_{1}(m,t)}{\partial t}=-\frac{\partial}{\partial m}\{(k_{m}-\frac{\gamma^{*}m}{m+\lambda}-\gamma_{m}\, m)\rho_{1}(m,t)\}+k_{a}\,\rho_{0}(m,t)-k_{d}\,\rho_{1}(m,t)\end{equation}The first terms in 
Eqs. (10) and (11) are the so-called `transport' terms, representing the net flow of the probability density.
 The second terms represent the gain/loss in the probability density due to random transitions between the state $j$ ($j=1,0$) and
 the other accessible state. In the steady state, both $\frac{\partial\rho_{0}}{\partial t}$ and $\frac{\partial\rho_{1}}{\partial t}$
are zero and the total probability density function $\rho$ satisfies
the equation

\begin{equation}
\frac{\partial\rho}{\partial m}=\frac{(\frac{k_{a}}{k_{m}}-f'(m))\rho}{f(m)}+\frac{(f'(m)-\frac{k_{d}}{k_{m}})\rho}{1-f(m)}\end{equation}where

\begin{equation}
\begin{array}{c}
f(m)=\frac{\gamma_{new}^{*}m}{m+\lambda}+\gamma_{new}m,\\
\gamma_{new}^{*}=\frac{\gamma*}{k_{m}},\:\:\gamma_{new}=\frac{\gamma_{m}}{k_{m}}\end{array}\end{equation}The steady state solution for $\rho$ is given by

\begin{equation}
\rho(m)=N\, m^{a_{1}}(d_{2}+m)^{a_{2}}(d_{3}+m)^{a_{3}}(d_{5}-m)^{a_{4}+a_{5}}(d_{6}+m)^{a_{4}-a_{5}}\end{equation}where N is the normalisation constant,

\begin{eqnarray}
a_{1}&=&\frac{k_{a}}{k_{m}}(\frac{\lambda}{\gamma_{new}^{*}+\gamma_{new}\lambda})-1 \nonumber \\
a_{2}&=&\frac{k_{a}}{k_{m}}(\frac{\gamma_{new}^{*}}{\gamma_{new}(\gamma_{new}^{*}+\gamma_{new}\lambda)})-1 \nonumber \\
a_{3}&=&2 \nonumber \\
a_{4}&=&\frac{k_{d}}{k_{m}}\frac{1}{2\gamma_{new}}-1 \nonumber \\
a_{5}&=&\frac{k_{d}}{k_{m}}(\frac{1-\gamma_{new}^{*}+\lambda\gamma_{new}}{2\gamma_{new}})\frac{1}{2\:\sqrt{\lambda\gamma_{new}+(\:\frac{\gamma_{new}^{*}+\lambda\gamma_{new}-1}{2}\:)^{2}}}  
\end{eqnarray}and

\begin{eqnarray}
d_{2}&=&\frac{\gamma_{new}^{*}}{\gamma_{new}}+\lambda \nonumber \\
d_{3}&=&\lambda \nonumber \\
d_{5}&=&-\frac{(\gamma_{new}^{*}+\lambda\gamma_{new}-1)}{2\gamma_{new}}+\frac{1}{2\gamma_{new}}\:\:\sqrt{(\gamma_{new}^{*}+\lambda\gamma_{new}-1)^{2}+4\lambda\gamma_{new}} \nonumber \\
d_{6}&=&\frac{(\gamma_{new}^{*}+\lambda\gamma_{new}-1)}{2\gamma_{new}}+\frac{1}{2\gamma_{new}}\:\:\sqrt{(\gamma_{new}^{*}+\lambda\gamma_{new}-1)^{2}+4\lambda\gamma_{new}}
\end{eqnarray}Putting $\gamma_{new}^{*}=0$ and $\lambda=0, $ i.e., considering only unregulated gene expression, one recovers from Eq. (14) the beta distribution
\cite{karma,raj}:

\begin{equation}
\rho(m)=N_{1} \; m^{(\frac{k_{a}}{\gamma_{m}}-1)}(\frac{k_{m}}{\gamma_{m}}-m)^{(\frac{k_{d}}{\gamma_{m}}-1)}\end{equation}where $N_{1}$ is the normalisation constant. In this case, binary gene expression is obtained in the parameter regime $r_{1}=\frac{k_{a}}{\gamma_{m}}$
and $r_{2}=\frac{k_{d}}{\gamma_{m}}$ both $<1$. Two prominent peaks in the probability density function appear when $r_{1}$ and $r_{2}$ are
 comparable in magnitude. In the case of transcription-factor regulated gene expression, the
effective activation and deactivation rate constants, $k_{a}^{'}(s)$
and $k_{d}^{'}(s)$, are functions of the regulatory protein (transcription
factor) concentration $s$. The gene expression response to a regulatory
stimulus may be either graded or binary. The response is quantified
in terms of the concentrations of mRNAs/proteins. In graded response,
the average steady state concentration of the gene expression product varies
continuously as the concentration $s$ of the regulatory molecules
is changed, until a saturation level is reached. In the case of binary
response, gene expression occurs at either of two average levels (say, low or high) and
expression at other levels is minimal. The fraction of cells
in the low/high expression level changes as $s$ is changed. This
gives rise to a bimodal distribution in the protein/mRNA levels in
an ensemble of cells. In the case of transcription-factor regulated
gene expression, binary gene expression is obtained when $r_{1}=\frac{k_{a}^{'}(s)}{\gamma_{m}}<1$ and $r_{2}=\frac{k_{d}^{'}(s)}{\gamma_{m}}<1$.
One should point out that in the parameter regime in which 
binary gene expression occurs, unimodal distributions are obtained when $r_{1}<<r_{2}$
or $r_{2}<<r_{1}$. The full parameter regime is associated with the system exhibiting binary response to changing activation and deactivation rate constants. 
For example, for $r_{2}>>r_{1}$, the probability distribution of mRNA levels has
a single peak at a low level. As $r_{1}$ increases, a second peak
appears at a high expression level with the peak becoming more prominent
as $r_{1}$ approaches $r_{2}$. At the other extreme of parameter
values, $r_{1}>>r_{2}$, a single peak at the high expression level
is obtained. As $r_{1}$ changes, the position of the peak remains the same, a characteristic of binary response. Stochastic binary gene expression refers to a bimodal
distribution of mRNA/protein levels and the bifurcation from a unimodal
to a bimodal distribution occurs in the parameter regime $r_{1}<1$
and $r_{2}<1$. 

In the case of miRNA-mediated regulation of gene expression, the most prominent contribution to binary gene expression is obtained
when both $a_{1}$ and $a_{4}+a_{5}$ are $<0$ in Eq. (14), i.e., when the following
inequalities are satisfied:

\begin{eqnarray}
\frac{k_{a}}{\gamma_{m}}&<&1+\frac{\gamma_{new}^{*}}{\lambda\gamma_{new}}\\ \nonumber
\frac{k_{d}}{\gamma_{m}}&<&1+\frac{1-f}{1+f}\quad,\quad f=\frac{\phi}{\sqrt{\phi{}^{2}+\lambda\gamma_{new}\gamma_{new}^{*}}} \nonumber
\end{eqnarray}with $\phi=\frac{(1-\gamma_{new}^{*}+\lambda\gamma_{new})}{2}$. One notes
that the parameter region in which stochastic binary gene expression
may be obtained is expanded from that $(\frac{k_{a}}{\gamma_{m}}<1$ and
$\frac{k_{d}}{\gamma_{m}}<1)$ in the case of unregulated gene expression. 
Figure 3 shows the steady state mRNA distribution  $\rho(m)$ versus $m$ when $a_{1}$ is $<0$ and $a_{4}-a_{5}<0$ for the 
parameter values (a) $r_{1}<1$, $r_{2}<1$ and (b) $r_{1}>1$, $r_{2}<1$. The other parameter values are $\gamma_{new}=0.1$ and $\gamma_{new}^{*}=0.2$. 
For each case, the values of $\lambda$ are $\lambda=0.001$, $\lambda=1$ and $\lambda=20$ respectively. We observe that increasing (decreasing) values of $\lambda$ 
disfavours (favours) bimodality. For $\gamma_{new}^{*}=0.2$, $\lambda=0$, i.e., the case of unregulated gene expression, binary gene expression occurs in the parameter 
regime $r_{1}<1$, $r_{2}<1$ but not in the regime $r_{1}>1$, $r_{2}<1$. Figure 3 demonstrates the enhanced possibility of binary gene expression in the 
case of miRNA-mediated regulation of gene expression. One notes that if the parameter $\lambda$ is set to zero in Eq. (16), the parameter regime in which binary gene expression 
is observed is expanded to include all values of the ratio $r_{1}=\frac{k_{a}}{\gamma_{m}}$ and $r_{2}=\frac{k_{d}}{\gamma_{m}}<1$. The parameter $\lambda$ ($=\frac{k_{off}+\gamma_{m^{*}}}{k_{on}}$) is 
analogous to the dissociation constant for the formation of the bound mRNA-miRNA complex with $\lambda=0$ signifying infinitely strong binding. We now provide a physical 
understanding of the enhanced occurrence of binary gene expression when a miRNA-regulated gene expression parameter, say $\lambda$, is changed.

As mentioned in the Introduction, molecular titration can generate ultrasensitive input-output responses. The origin of ultrasensitivity lies in the sequestration 
of an active component in an inactive complex through binding to an antagonist \cite {mukherji, buchler, bet}. The simple mechanism involves the kinetic scheme

\begin{figure}
\centering{}
\includegraphics[scale=0.7]{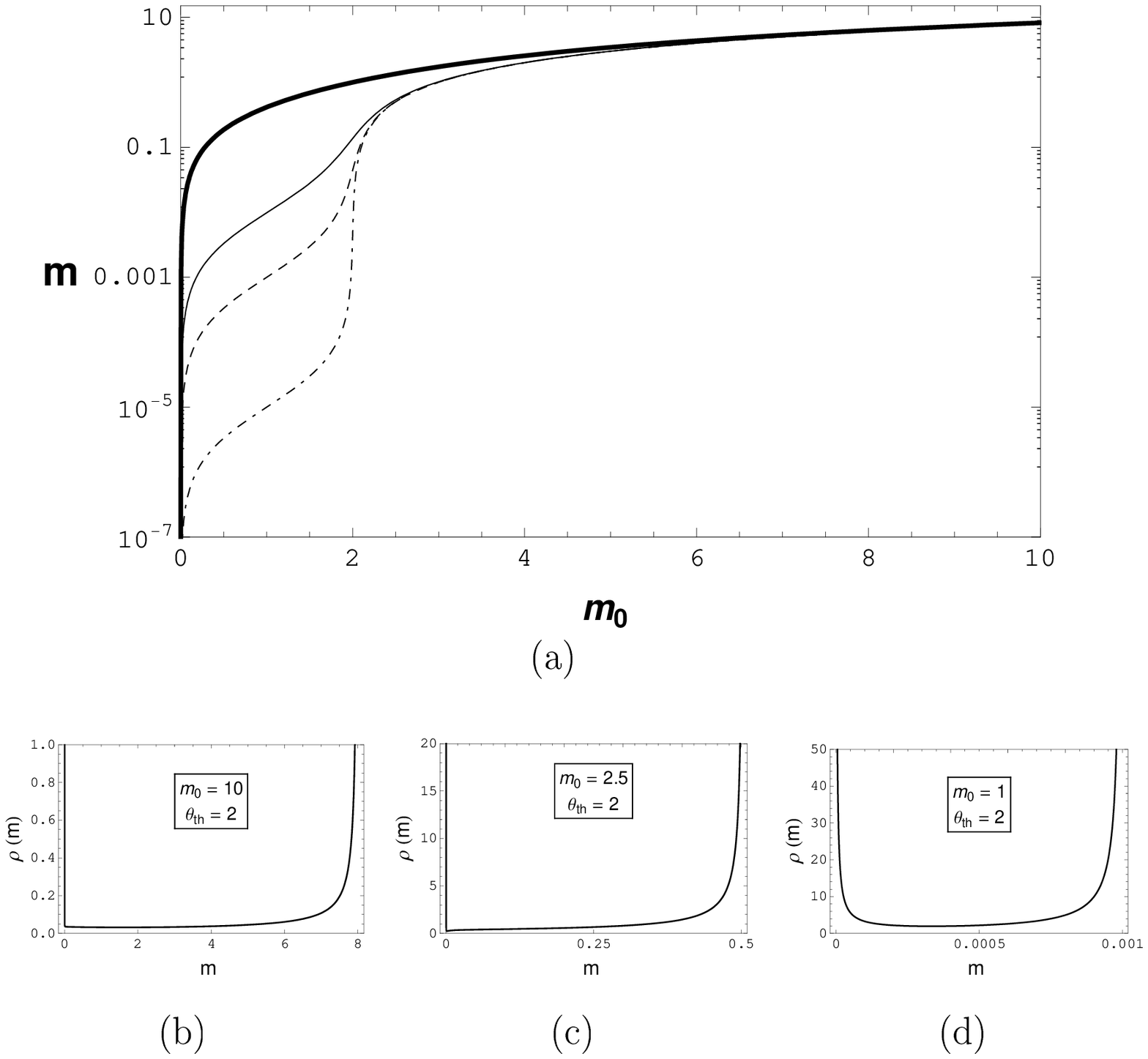}
\caption{(a) Steady state concentration of mRNA $m$ versus total mRNA concentration $m_{0}$ (Eq. (20)). The parameter $\theta_{th}$ has the value 2 with
 $\lambda=10^{-5}$ (dash-dotted line), 0.001 (dashed line), 0.01 (thin solid line) and 0.1 (thick solid line). Marked ultrasensitive response 
is obtained for very low values of $\lambda$. In (b), (c) and (d), the $\rho\,(m)$ versus $m$ plots are shown for $m_{0}>\theta_{th}$ (b), $m_{0}\simeq\theta_{th}$ (c) and 
$m_{0}<\theta_{th}$ (d). In the last case, the second peak position is at a value very close to zero.}
\label{Fig_ultra}
\end{figure}

\begin{equation}
X+Y \rightleftarrows XY
\end{equation}where X represents the active molecule, Y its antagonist and XY the inactive complex. Let $K_{d}$ be the dissociation constant associated with the complex XY. The total 
concentrations of X and Y are given by $X_{T}=[X]+[XY]$ and $Y_{T}=[Y]+[XY]$. When $Y_{T}>>K_{d}$, the X versus $X_{T}$ plot exhibits ultrasensitivity with the threshold 
set by $Y_{T}$. For $X_{T}<Y_{T}$, almost all the X molecules are part of the inactive complex $XY$ so that the concentration of free X molecules is very low. At 
the threshold point $X_{T} \simeq Y_{T}$, the sequestration of X molecules is no longer dominant, so that a small increase in $X_{T}$ can give rise to a large increase in X. 
The sharpness of the ultrasensitive response in the simple example is controlled by the ratio $\frac{Y_{T}}{K_{d}}$ \cite {buchler}. In the present study, X and Y represent the 
mRNA and the miRNA respectively. When $a_{4}+a_{5}$ is $<0$ in Eq. (14), a dominant singularity occurs at the point $m=d_{5}$, with the expression for $d_{5}$ given in 
Eq. (16). In terms of the original parameters, $d_{5}$ is given by

\begin{equation}
 d_{5}= \frac {1}{2} [(m_{0}-\lambda - \theta_{th}) + [(m_{0}-\lambda - \theta_{th})^{2} + 4m_{0} \lambda]^{1/2}]
\end{equation}
 where $m_{0}=\frac {k_{m}}{\gamma_{m}}$, $\theta_{th}=\frac {\gamma^{*}}{\gamma_{m}}=\frac {\gamma_{m}^{*} mi_{T}}{\gamma_{m}}$ and $\lambda$ is given by the expression in Eq. (6). 
The magnitude of $m_{0}$ is a measure of the total amount of mRNA.
When the effective dissociation constant $\lambda\rightarrow0$, we
obtain

\begin{eqnarray}
m=d_{5} &=& \frac{1}{2}\,\,[m_{0}-\theta_{th}+\sqrt{(m_{0}-\theta_{th})^{2}}]\\
&=& 0 \: \: \: \mbox{if} \: \: \: m_{0}<\theta_{th}\\
&=& m_{0} \: \: \: \mbox{if} \: \: \:  m_{0}>\theta_{th}
\end{eqnarray}The constant $\theta_{th}$ is proportional to the total amount, $mi_{T}$,
of miRNAs and sets the threshold of an ultrasensitive steady state
response. Figure 4 (a) shows the steady state $m$ versus $m_{0}$ plots (Eq. (20))
on a semi-logarithmic scale for the values of $\lambda=10^{-5}$ (dash-dotted
line), $\lambda=0.001$ (dashed line), $\lambda=0.01$ (thin solid
line) and $\lambda=0.1$ (thick solid line). One finds that the ultrasensitive
behaviour is prominent only when $\lambda$ is close to zero. The
value of $\theta_{th}$ is fixed at the value 2. Figures 4 (b)-(d)
show the steady state $\rho\,(m)$ versus $m$ curves (Eq. (14))
for the parameters values $\gamma_{new}^{*}=0.2$, $\gamma_{new}=0.1$ and $m_{0}$ greater than $\theta_{th}$ (b), almost
equal to $\theta_{th}$ (c) and less than $\theta_{th}$ (d). In the last case, the
second peak at $m=d_{5}$ ($a_{4}+a_{5}<0$ in Eq. (14)) occurs
very close to the peak at $m=0$ ($a_{1}<0$).

\noindent The inequalities in Eq. (18) can be rewritten in a simpler form

\begin{equation}
\frac{k_{a}}{\gamma_{eff1}}<1\,\,\,\,\,,\,\,\,\,\frac{k_{d}}{\gamma_{eff2}}<1
\end{equation}with $\gamma_{eff1}=\frac{\gamma*}{\lambda}+\gamma_{m}$ and $\gamma_{eff2}=\gamma_{m}(1+\frac{1-f}{1+f})$.

\noindent The rate constants $\gamma_{eff1}$ and $\gamma_{eff2}$ can be interpreted
as effective mRNA decay rate constants by noting the following. From
Eq. (9), for $z=0$, i.e., when the target gene is in the inactive state,
\begin{equation} 
\frac{dm}{dt}=-\frac{\gamma^{*}\, m}{m+\lambda}-\gamma_{m\,}m=f(m) \nonumber
\end{equation}On linearising $f\:(m)$ around the steady state value $m=0$,
one obtains

\begin{equation}
\frac{dm}{dt}= -(\frac{\gamma^{*}}{\lambda}+m)\:m = -\gamma_{eff1}\:m
\end{equation}Similarly, for $z=1$, i.e., when the target gene is in the active
state, $f\:(m)=k_{m}-\frac{\gamma^{*}\, m}{m+\lambda}-\gamma_{m\,}m$.
Again, linearising $f\:(m)$ around the steady state value $m=d_{5}$,
one gets 

\begin{equation}
\frac{dm}{dt}=k_{m}-\gamma_{m\,}(1+\frac{1-f}{1+f})\: m=k_{m}-\gamma_{eff2}\:m
\end{equation}Eqs. (23) and (24) are identical in forms to the equations obtained
in the case of unregulated gene expression. In the latter case, however,
$\gamma_{eff1}=\gamma_{eff2}=\gamma_{m}$. The origin of binary gene
expression in the parameter regime $\frac{k_{a}}{\gamma_{eff1}}<1$,
$\frac{k_{d}}{\gamma_{eff2}}<1$ now has a clear physical interpretation.
In the inactive state of the gene, the activation rate constant has
a lower value than that of the effective mRNA degradation rate constant $\gamma_{eff1}$
so that the accumulated protein level can decay to the steady state
level $m=0$ before the next transition to the active state of the
gene occurs. Similarly, in the active state of the gene, the deactivation
rate constant has a magnitude lower than that of the effective mRNA
degradation rate constant $\gamma_{eff2}$. The gene is thus in the
active state for a sufficiently long time so that the steady state
mRNA level $m=d_{5}$ (Eq. (20)) is attained. The parameter regime
(Eq. (22)) corresponds to binary response in gene expression
as changing the $k_{a}$, $k_{d}$ values does not alter the peak
positions but only the amplitudes. When both $\frac{k_{a}}{\gamma_{eff1}}$
and $\frac{k_{d}}{\gamma_{eff2}}$ are $>1$, the activation/deactivation
rate constant is larger than the effective mRNA degradation rate constant, so that 
 a transition from the inactive to the active state and
vice versa occurs before the mRNA level can attain the value $m=0$
in the inactive state, and the value $m=d_{5}$ in the active state.
As a result, the probability distribution $\rho\,(m)$ is unimodal, with the
peak position at an intermediate mRNA level. One also obtains a graded
response to changing values of the rate constants $k_{a}$ and $k_{d}$,
i.e., the position of the peak shifts in a graded manner. In the steady
state expression for $\rho\,(m)$, for $a_{1}<0$ and $a_{4}+a_{5}<0$,
the singularities occur at $m=0$ and $m=d_{5}$. The other terms
add a small background correction to the distribution. Figure 5 shows
the plot of $\rho\,(m)$ versus $m$ in the case when $a_{1}<0$ and
$a_{4}-a_{5}<0$, ($a_{4}+a_{5}>0$). In this case, though a bimodal
distribution is obtained for $r_{1}$, $r_{2}>1$ and $\lambda$ small,
the second peak is associated with quite a broad distribution ((a)
and (b)). When the magnitude of $\lambda$ is increased from $\lambda=0.002$
(b) to $\lambda=0.2$ (c), a unimodal distribution is obtained.

\begin{figure}
\centering{}
\includegraphics[scale=0.85]{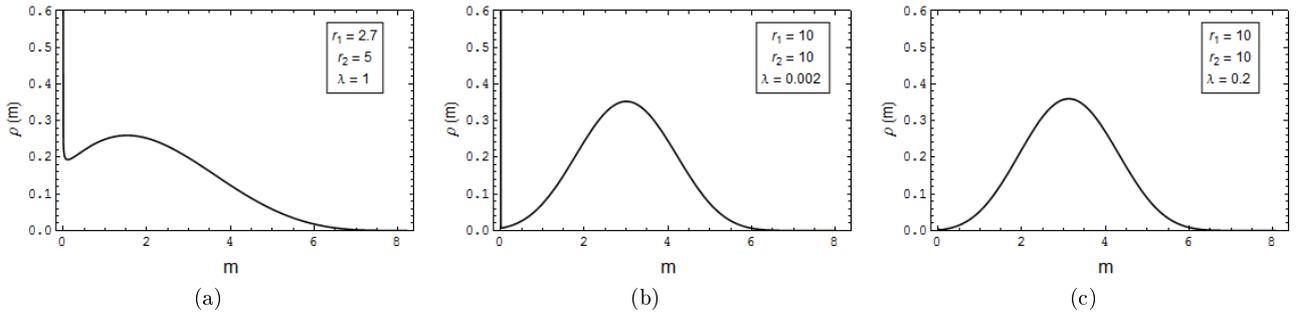}
\caption{ $\rho\,(m)$ versus $m$ plots in the steady state for $a_{1}< 0$ and $a_{4}-a_{5} < 0$. The values of the parameters are 
quoted in individual boxes.  The other parameter values are $\gamma_{new} = 0.1$, $\gamma_{new}^{*} = 0.2$. For low values of $\lambda$,  a 
bimodal distribution is obtained ((a) and (b)) with the second peak more prominent for the lower value of $\lambda$. When  $\lambda$ 
is increased from the value 0.002 in (b) to the value 0.2 in (c), the bimodality in destroyed and one gets a unimodal distribution.}
\label{Fig_a4-a5}
\end{figure}

\section{Conclusion}

Theoretical modeling studies combined with experiments have been spectacularly
 successful in uncovering novel features of cellular phenomena. One such feature
 is that of binary gene expression, in which the distribution of mRNA/protein levels
 has two prominent peaks. The possible origins of binary gene expression are several of which
 three principal mechanisms have been the focus of recent studies \cite{bose}:
 positive feedback-based \cite{pomer, ferel}, emergent bistability \cite{tan, ghosh} and purely stochastic
 \cite{karma,ko, kepler, kaern,raj2, karma2, to}. The first two mechanisms create the potential for the coexistence
 of two stable expression states and noise-induced transitions between the states give
 rise to the bimodal nature of the distribution of mRNA/protein levels. In the case of
 binary gene expression with a purely stochastic origin, there is no bistability in the
 deterministic limit. Experimental observations of stochastic binary gene expression
 have been reported in a number of studies \cite{kaern, raj2, to}, in agreement with theoretical
 results. The large number of studies carried out so far on binary gene expression
 consider the expression to be unregulated or transcription-factor regulated. The
 issue of binary gene expression in the case of post-transcriptional regulation of
 gene expression has been mostly confined to the investigation of models in which
 both transcriptional and post-transcriptional (e.g., miRNA-regulated) modes of
 regulation are considered and the transcriptional regulation involves a cooperative
 positive feedback.
In this paper, we have discussed two possible scenarios for miRNA-mediated
regulation of gene expression, one deterministic and the other stochastic,
and demonstrated the existence of binary gene expression in each case.
In the first case, a nonlinear protein decay term, together with a noncooperative
positive feedback term generate bistability in specific parameter regimes.
In the second case, the conditions for obtaining stochastic binary
gene expression are obtained and one finds that the parameter region in which
 binary gene expression occurs is more extended in comparison with the cases
 of unregulated and transcription factor-regulated gene expression. In the latter
 two cases, slow transitions between the inactive and
active gene expression states are responsible for binary gene expression.
The conditions $\frac{k_{a}}{\gamma_{m}}<1$ and $\frac{k_{d}}{\gamma_{m}}<1$
can be reexpressed as $\frac{T_{a}}{T_{m}}>1$ and $\frac{T_{d}}{T_{m}}>1$
with $T_{a}$, $T_{d}$ being the average lifetimes of the active
and inactive gene expression states ($T_{a}\sim\frac{1}{k_{a}}$,
$T_{d}\sim\frac{1}{k_{d}}$) and $T_{m}$ is the average lifetime of the 
mRNAs. In the case of miRNA-mediated regulation of gene expression,
ultrasensitivity generated by molecular titration plays a key role
in the generation of binary gene expression. The conditions $\frac{k_{a}}{\gamma_{eff1}}<1$
and $\frac{k_{d}}{\gamma_{eff2}}<1$ (Eq. (22)) set new time
scales in the characterisation of the `slowness' of the transitions between the gene
expression states. This results in an expanded parameter regime
in which binary gene expression is observed. The theoretical predictions
made in the present study could be tested in experiments involving
natural and synthetic gene circuits.

Binary gene expression generates phenotypic heterogeneity in a cell
population with identical genetic makeup and exposed to the same initial
conditions and environment. Microorganisms adopt a number of strategies for
coping with stressful situations like environmental fluctuations,
nutrient depletion and application of antibiotic drugs. One such strategy
is the creation of phenotypic heterogeneity so that the whole population does
not suffer the same fate when subjected to stress \cite{balazsi, balaban, bet}.
Recent experiments on \emph{E. coli }\cite{balazsi, balaban}, \emph{B. Subtilis} \cite{bet}
and \emph{M. Smegmatis} \cite{plos,BMC} have demonstrated the advantages of phenotypic
heterogeneity when microorganisms are subjected to stress. The regulation
of gene expression by miRNAs is known to be activated in a number
of cases in which a cell population is subjected to stress \cite{inui, flynt, leung}.
Experimental evidence of phenotypic heterogeneity in the form of binary
gene expression due to regulation by miRNAs would be of significant
interest in the context of the response of cell populations to stressful
conditions.

\begin{center}\textbf{Acknowledgment}\end{center}

\noindent SG acknowledges the support by CSIR, India, under Grant No.  09/015(0361)/2009-EMR-I.

\end{document}